# Honeycomb oxide heterostructure: a new platform for Kitaev quantum spin liquid

Baekjune Kang,[1†] Miju Park,[1‡] Sehwan Song,[2] Seunghyun Noh,[3] Daeseong Choe,[3] Minsik Kong,[2] Minjae Kim,[2] Choongwon Seo,[1] Eun Kyo Ko,[4,5] Gangsan Yi,[1] Jung-woo Yoo,[3] Sungkyun Park,[2] Jong Mok Ok,[2**] and Changhee Sohn[1*]

[1]*Department of Physics, Ulsan National Institute of Science and Technology, Ulsan, 44919, Republic of Korea*

[2]*Department of Physics, Pusan National University, Pusan, 46241, Republic of Korea*

[3]*Department of Materials Science and Engineering, Ulsan National Institute of Science and Technology, Ulsan, 44919, Republic of Korea*

[4]*Center for Correlated Electron Systems, Institute for Basic Science (IBS), Seoul, 08826, Republic of Korea*

[5]*Department of Physics and Astronomy, Seoul National University, Seoul, 08826, Republic of Korea*

**Abstract**

Kitaev quantum spin liquid, massively quantum entangled states, is so scarce in nature that searching for new candidate systems remains a great challenge. Honeycomb heterostructure could be a promising route to realize and utilize such an exotic quantum phase by providing additional controllability of Hamiltonian and device compatibility, respectively. Here, we provide epitaxial honeycomb oxide thin film $Na_3Co_2SbO_6$, a candidate of Kitaev quantum spin liquid proposed recently. We found a spin glass and antiferromagnetic ground states depending on Na stoichiometry, signifying not only the importance of Na vacancy control but also strong frustration in $Na_3Co_2SbO_6$. Despite its classical ground state, the field-dependent magnetic susceptibility shows remarkable scaling collapse with a single critical exponent, which can be interpreted as evidence of quantum criticality. Its electronic ground state and derived spin Hamiltonian from spectroscopies are consistent with the predicted Kitaev model. Our work provides a unique route to the realization and utilization of Kitaev quantum spin liquid.

**Main**

Quantum entanglement, a subtle nonlocality that Einstein never liked, becomes a central subject again in the wide range of science and technology[1-4]. Quantum entangled states in many-body condensed matter are particularly interesting as they possess nontrivial topology and fractional excitations, which contributes to the robustness of entanglement against external perturbation[5]. Kitaev quantum spin liquid (QSL) is a notable example. In this model, spins in a honeycomb lattice are frustrated by the bond-directional Ising interaction, resulting in a massively entangled quantum spin state with fractional Majorana fermions and non-Abelian anyons[6]. As other types of QSL[7-9], such an exotic phase, however, is so scarce that searching for new candidate materials remains a great challenging task[10-19].

Being believed to be impossible and barely conducted[20], heterostructure approaches could be a promising route to realizing and utilizing Kitaev QSL. Every proposed material for Kitaev QSL has classical antiferromagnetic ground states due to the presence of non-Kitaev interaction originating from additional structure distortions or long-range spin interaction[10,15,21]. However, all the existing experiment on Kitaev QSL has been performed in a bulk system, where control of spin Hamiltonian is limited. In heterostructure geometry, on the other hand, symmetry, strain, and interfacial engineering can be utilized for extensive manipulation of spin Hamiltonian[22,23]. In addition, we have witnessed the triumphs of heterostructure approaches in semiconductor and spintronic technologies[24,25]. Therefore, we can expect that similar heterostructure approaches in quantum magnetism can advance breakthroughs in quantum information technology like topological quantum computation[5].

To utilize heterostructure for Kitaev QSL, however, two major difficulties must be carefully handled. First, most experiment for QSL is bulk-sensitive including inelastic neutron scattering[14] and thermal Hall experiments[17], which can be overcome, however, by performing the newly proposed experiment based on non-local spin transport[26]. Second, it remains elusive whether Kitaev QSL candidate materials can be synthesized in heterostructures with minimal disorders[27,28]. There have been *zero* studies about the synthesis of Kitaev QSL candidates in heterostructures, with any signs of quantum criticality. In addition, even bulk compounds suffer from disorders, which hinder obtaining intrinsic magnetic and quantum properties of systems[21,29]. It is a common belief that heterostructures are not a suitable platform for quantum magnetism due to more impurities than in bulk.

In this letter, we provide a synthesis of honeycomb oxide heterostructures $Na_3Co_2SbO_6$, a candidate material for Kitaev QSL[18,21,30-35], and experimental signatures of quantum criticality. We have successfully synthesized stoichiometric, single-phase, and epitaxial $Na_3Co_2SbO_6$ thin films on ZnO substrates. We demonstrate the precise control of Na stoichiometry of the thin film via growth oxygen partial pressure. As we minimized Na vacancies, our thin film shows long-range antiferromagnetic ordering near 8.3 K, the same as the highest reported value in stoichiometric bulk samples. The field-dependent magnetic susceptibility shows remarkable scaling collapse as a function of single variable $T/H$, a signature of quantum criticality. We further conducted optical and ultraviolet photoemission spectroscopy to reveal the quantitative parameters of the electronic Hamiltonian of the system including charge transfer energy ($\Delta_{pd}$), crystal field splitting ($10Dq$), Hund's coupling ($J_H$), and Coulomb interaction ($U$). A constructed spin Hamiltonian from those parameters is consistent with the proposed Kitaev model.

Among the candidate materials for Kitaev QSL, $Na_3Co_2SbO_6$ could be a promising system for heterostructure approaches as follow. The crystal structure of $Na_3Co_2SbO_6$ is monoclinic layered oxides, where $Na^+$ layers and $(Co_{2/3}Sb_{1/3}O_2)^-$ layers are alternatively stacked (Figure S1). The $Co^{2+}$ ions in the cobalt-antimony layer construct the honeycomb lattice. Although $Na_3Co_2SbO_6$ has a smaller degree of spin-orbit coupling compared to other candidates like $\alpha$-$RuCl_3$ and $Na_2IrO_3$, the $J_{eff}= 1/2$ state appears as a ground state due to the unquenched orbital momentum of the $Co^{2+}$ [18,33,34]. In addition, the localized $3d$ orbitals and charge-transfer insulating state is favored over extended $4d$/$5d$ orbitals and Mott insulating state in Kitaev QSL. The localized $3d$ orbital suppresses the undesired next-nearest neighbor exchange interaction and the charge-transfer nature prefers superexchange via oxygen which vanishes the isotropic exchange term in the 90° bonding geometry rather than a direct exchange between cobalt ions[36]. Several experiments on bulk $Na_3Co_2SbO_6$ have shown convincing evidence about the presence of strong frustration and large Kitaev exchange terms[18,37]. In particular, it has been theoretically proposed that $Na_3Co_2SbO_6$ can have the Kitaev QSL as a ground state by simply controlling trigonal distortion[34], thereby making $Na_3Co_2SbO_6$ a suitable model system for heterostructure approaches.

The hexagonal substrate ZnO along the [0001] direction provides a good epitaxial relationship with $Na_3Co_2SbO_6$ along the [001] direction, promoting the synthesis of a high-quality thin film. As schematically shown in Figure 1a, the hexagonal substrate ZnO (0001) surface exhibits 2D six-fold symmetry with coincident oxygen positions with $Na_3Co_2SbO_6$. Because of similarities in the crystal structure of the $Na_3Co_2SbO_6$ (001) plane and the ZnO (0001) plane, we expect the growth of epitaxial thin films despite the large lattice mismatch between them (~4.2 %). Figure 1b is the X-ray diffraction (XRD) $\theta$-$2\theta$ scan of $Na_3Co_2SbO_6$ 60 nm thickness thin film grown on ZnO [0001] substrate by pulsed laser deposition. We observe 00$l$ diffraction peaks of $Na_3Co_2SbO_6$ consistent with reported values of single crystals. Figure 1c displays a rocking curve of the $Na_3Co_2SbO_6$ 001 peak. It is fitted by two Lorentzian functions, typically found in relaxed films. The full width at half maximum (0.055°) of the narrower peak verifies high crystal quality.

We demonstrate that the sodium stoichiometry of $Na_3Co_2SbO_6$ can be precisely controlled by adjusting the background oxygen partial pressure ($P_{O_2}$) during the thin film deposition process. Note that, control of sodium stoichiometry has been known to be difficult even in single crystals due to the high reactivity and volatility of sodium ions[21]. As a result, Neel temperature $T_N$ of powder and single crystals $Na_3Co_2SbO_6$ varies greatly from 4.4 to 8.3 K with different sodium stoichiometry[21,30-32,35]. In the thin-film deposition process, however, we can control the sodium stoichiometry by changing $P_{O_2}$. According to the previous study on $LiCoO_2$[38], the number of light elements like Li and Na ions reaching the substrate decreases as the number of collisions with oxygen increases. Figure 2a is the $\theta$-$2\theta$ scan near 001 peak of $Na_3Co_2SbO_6$ of different $P_{O_2}$. As $P_{O_2}$ is increased, the 001 peak shifts to a low angle indicating $c$-axis elongation. It is consistent with the previous report on $Na_xCoO_2$[39], where the $c$-axis elongated as decreasing $x$. Figure 2b shows the lattice constants of $c^*$ as a function of different $P_{O_2}$. We found $P_{O_2}$ below 1 mTorr is required to minimize Na vacancies in heterostructures.

The magnetic ground states of $Na_3Co_2SbO_6$ evolve from antiferromagnetic to spin-glass states with Na vacancies, indicating not only the importance of Na stoichiometry control but also the existence of strong frustrations in our heterostructures. Figure 2c is the zero-field cooling (ZFC) and field cooling (FC) magnetic susceptibility $\chi(T)$ measured along the Co-Co bonding direction. The $\chi(T)$ of the Na-stoichiometric sample grown at $P_{O_2} = 0$ mTorr shows a distinct anomaly at 8.3 K with a negligible bifurcation of ZFC and FC, indicating antiferromagnetic ground states. Note that this temperature is the same with the highest reported value of bulk compounds[21,35]. In Na-vacant thin film grown at high $P_{O_2}$, on the other hand, a kink from antiferromagnetic ordering and bifurcation between FC and ZFC becomes weak and conspicuous, respectively, resulting in spin glass states. This observation indicates the control of Na vacancy is essential to identify the true magnetic ground state of $Na_3Co_2SbO_6$. In addition, the emergence of spin glass behavior with small off-stoichiometry implies the presence of strong frustration, a key ingredient for quantum magnetism. The previous theoretical calculation suggests the existence of a subtle balance between ferromagnetic and antiferromagnetic exchange interactions, consistent with the emergence of spin glass states with minimal Na vacancies[34].

The antiferromagnetic ground state in $Na_3Co_2SbO_6$ films is very fragile against an external magnetic field, another evidence that the system is close to the quantum critical region. Figure 3a is in-plane $\chi(T)$ with different applied magnetic fields $H$. The evident kink at $H = 1$ kOe smeared out as the magnetic field raise over the critical magnetic field around 10 kOe. That suppression of the antiferromagnetic ground state has been observed also in $\alpha$-$RuCl_3$, a promising candidate of Kitaev QSL. However, the critical field $H_c$ of $Na_3Co_2SbO_6$ is substantially smaller than that of $\alpha$-$RuCl_3$ despite similar $T_N$, implying that the system locates closer at the borderlines between ferromagnetic and antiferromagnetic phases. The suppression of antiferromagnetic ordering observed in our heterostructure is also consistent with recent inelastic neutron scattering on high-quality single crystals that antiferromagnetic ground states evolve into ferrimagnetic and fully polarized states with an applied magnetic field[35].

Despite classical ground states of $Na_3Co_2SbO_6$, we found a remarkable scaling collapse of $\chi(T, H)$ as a function of single variable $T/H$ in all field regions, a signature of quantum criticality. According to previous studies on $\alpha$-$RuCl_3$, strong quantum fluctuation can be captured in high-temperature phases where thermal fluctuation destroys classical ground states[40,41]. Figure 3b shows the scaling collapse of $H^{0.86}\chi(T, H)$ as a function $T/H$. $H^{0.86}\chi(T, H)$ converges to a single line in the range 10-50 K where not only Curie-Weiss law is deviated[32] but also the classical ground state is destroyed by thermal fluctuations. Remarkably, all the $\chi(T, H)$ collapse into a single curve with a single critical exponent and single variable. The contribution to the magnetization of disordered local moments in a frustrated quantum spin system is expressed as $M[H, T] \sim HT^{-\gamma}$ in the range $H \ll T$ [42]. As a result, the relation $\log(H^{-\gamma}\chi(H, T)) \sim -\gamma\log(T/H)$ is satisfied and a linear function emerges, where the slope indicates critical exponent $-\gamma \sim 0.86$. The scaling collapse in all magnetic fields indicates the robustness of quantum criticality in $Na_3Co_2SbO_6$.

We further conducted optical spectroscopy on $Na_3Co_2SbO_6$ to extract quantitative parameters for its electronic and spin Hamiltonian. Figure 4a is the real part of optical conductivity $\sigma_1(\omega)$ of the $Na_3Co_2SbO_6$ thin film on the $MgAl_2O_4$ substrate. Due to the sharp absorption of the ZnO substrate, it is difficult to remove the artificial kink of $Na_3Co_2SbO_6$ on ZnO (figure S6). The peak position of $Na_3Co_2SbO_6$ on the $MgAl_2O_4$ substrate is the same as in the ZnO case. It can be explained by three interband transitions labeled $\alpha$, $\beta$, and $\gamma$ in the order of increasing energy. We obtained transition energies of those peaks with Lorentz oscillator models; the energies of the three peaks are 4.5eV ($\alpha$), 5.4 eV ($\beta$), and 6.45 eV ($\gamma$). We regard all Lorentz oscillators as interband transitions between the oxygen 2$p$ band to the cobalt 3$d$ band ($p$-$d$ transition). Due to the high electrical repulsion among seven electrons in the localized 3$d$ band, the $U$ of the Co$^{2+}$ ion is generally higher than the $\Delta_{pd}$ [43].

We extract the quantitative parameters for electronic structures of Na$_3$Co$_2$SbO$_6$ from the multi-orbital Hubbard model. The inset of Figure 4a represents the three possible $p$-$d$ transitions in the high spin 3$d^7$ system (Co$^{2+}$). The lowest transition is from O 2$p$ to Co $t_{2g}$ orbitals ($^3$A$_2$ symmetry), whose energy is defined by $\Delta_{pd}$. The other two transitions are from O 2$p$ to Co $e_g$ orbitals. Because of different orbital overlaps in the final state ($^3$T$_1$ and $^3$T$_2$ symmetry), each excitation requires different energies, $\Delta_{pd}$ +10$Dq$ ($^3$T$_2$) and $\Delta_{pd}$ +10$Dq$+3/2$J_H$ ($^3$T$_1$) [44]. Based on the multi-orbital Hubbard model, we assigned peaks α, β, and γ as $^3$A$_2$, $^3$T$_1$, and $^3$T$_2$ transitions. From the above equations, we calculate the quantitative value of three electronic parameters, $\Delta_{pd}$, 10$Dq$, and $J_H$ to 4.5 eV, 0.9 eV, and 0.7 eV, respectively.

To estimate $U$, we performed ultraviolet photoelectron spectroscopy. Figure 4b exhibits the photoelectron spectrum of the Na$_3$Co$_2$SbO$_6$ film with two distinct peaks δ and ε. Following the previous studies on CoO[45], we assign peaks δ and ε to transition $d^7$ → $d^7\underline{L}$, and $d^7$ → $d^6$ excitation, respectively. The energy difference between the two peaks corresponds to $U$ - $\Delta_{pd}$. These difference of Na$_3$Co$_2$SbO$_6$ has the same value as that of CoO, but the charge-transfer energy is about 1eV smaller. We estimate the $U$ ~ 8 ~ 10 eV considering the hybridization shift suggested in the previous study[45].

We reconstruct the spin Hamiltonian of Na$_3$Co$_2$SbO$_6$ based on electronic structure parameters by following the previous theoretical spin model[33]. The spin-exchange Hamiltonian of Na$_3$Co$_2$SbO$_6$ without trigonal distortion is written as

$$H = J\sum_{<ij>} S_i \cdot S_j + K\sum_{<ij>,a=b} S_i^a S_j^b + \Gamma \sum_{<ij>,a\neq b} S_i^a S_j^b$$

where *J*, *K*, and *Γ* are isotropic Heisenberg coefficient, anisotropic Kitaev coefficient, and off-diagonal term, respectively. The *J*, *K*, and *Γ* are the function of the electronic parameters, $\Delta_{pd}$, $10Dq$, $J_H$, and *U*. Due to $\Gamma \sim 0.001\,K$ in the $U > 5$ eV range, we only calculate *J* and *K* based on the spin-exchange interaction model of the $3d^7$ cobalt compound without trigonal distortion. Figure 4c exhibits the *J* and *K* as a function of *U* with a unit of $t^2/U$. We follow the calculation in the previous report with ($\Delta_{pd}$, $10Dq$, $J_H$) = (4.5 eV, 0.9 eV, 0.7 eV) obtained above[33]. We found a small *J* in the range of *U* between 8 ~ 10 eV (shaded region). Therefore, our spectroscopic approaches imply that the spin Hamiltonian of $Na_3Co_2SbO_6$ heterostructures can be possibly described by Kitaev physics if the trigonal distortion is suppressed.

In summary, we provided epitaxial honeycomb oxide heterostructure $Na_3Co_2SbO_6$ with signatures of quantum criticality. Its electronic and magnetic properties can be explained in terms of proposed Kitaev physics. The controllability and device compatibilities heterostructure provides could be a promising route to the future realization and application of the Kitaev phases. For example, providing $Na_3Co_2SbO_6$ heterostructures will enable us to engineer the trigonal crystal field, the major source of the antiferromagnetic ordering in $Na_3Co_2SbO_6$. Using isotropic substrates could also help to reduce the orthorhombic distortions existing in bulk $Na_3Co_2SbO_6$ as well[35]. Besides Kitaev physics, honeycomb oxide heterostructures have been proposed for a variety of correlated topological phases[46]. The demonstration of the successful growth of honeycomb oxides, therefore, will promote further searching for topological insulators and quantum anomalous Hall insulators in strongly correlated electron systems.

**Methods**

Sample preparation

Pulsed laser deposition is used to synthesize high-quality $Na_3Co_2SbO_6$ thin films. The O-faced ZnO [0001] and $MgAl_2O_4$ substrates (product of Crystal. Co) were annealed for 1 hour at 1100 °C and 6 hours at 1350 °C in the ambient pressure, respectively. The $Na_3Co_2SbO_6$ thin film growth on the treated substrates was performed in a wide range of growth parameters, such as temperature $T = 450 \sim 800$ °C, oxygen partial pressure $P_{O_2}= 0 \sim 2\times10^{-1}$ Torr, the energy of the KrF excimer laser ($\lambda = 248\ nm$) $E = 0.9 \sim 2$ J/cm$^2$, and laser repetition rate 2 ~ 10 Hz. The chamber base pressure was maintained under the $1\times10^{-6}$ Torr. The optimized growth condition on ZnO was $T = 625$ °C, $P_{O_2} = 0$ mTorr, $E = 0.9$ J/cm$^2$, and laser repetition = 10 Hz while optimized growth condition on $MgAl_2O_4$ and $T = 625$ °C, $P_{O_2} = 2$ mTorr, $E = 0.9$ J/cm$^2$, and laser repetition = 10 Hz. All the films for the experiments had a thickness of about 60 nm.

Characterization of $Na_3Co_2SbO_6$ Thin film

We used a D8 Discovery High-Resolution X-Ray Diffraction (Bruker) with Lynxeye detector to obtain $\theta$-$2\theta$ and the rocking curve of $Na_3Co_2SbO_6$ thin film. A superconducting quantum interference device (Quantum Design) is employed to investigate the magnetic properties of $Na_3Co_2SbO_6$ thin films and bare substrate ZnO. The magnetic moment of $Na_3Co_2SbO_6$ is extracted by deducting mass-normalized substrate data.

Optical and ultraviolet photoemission spectroscopy of $Na_3Co_2SbO_6$ Thin film

By utilizing an M-2000 ellipsometer (J. A. Woolam Co.), two ellipsometry parameters of $Na_3Co_2SbO_6$ thin film and a bare substrate ZnO and $MgAl_2O_4$ were independently measured at room temperature. A two-layer model was used for determining the optical constant of the $Na_3Co_2SbO_6$ thin film. To obtain the electronic structure of valance bands, we conducted ultraviolet photoemission by utilizing ESCALAB 250 XI (Thermo Fisher Scientific). For ultraviolet photoemission, the He I (22.1 eV) energy source is used. The Fermi level of $Na_3Co_2SbO_6$ is estimated from the Fermi level of reference gold.

**Acknowledgement**

Gangsan Yi was a talented, enthusiastic student and will be sadly missed. We thank Eun-Gook Moon and Jaehoon Kim for the fruitful discussion. This work is mainly supported by the National Research Foundation (NRF) funded by the Ministry of Science and ICT(2020R1C1C1008734) and Creative Materials Discovery Program through the National Research Foundation of Korea (NRF) funded by the Korea government (MSIT) (2017M3D1A1040834). J.O. acknowledges from Korea Basic Science Institute (National research Facilities and Equipment Center) grant funded by the Ministry of Education (Grant No. 2021R1A6C101A429) and National Research Foundation of Korea (NRF) grant funded by the Korea government (MSIT) (No. 2021R1F1A1056934). E.K.K is supported by Research Center Program of the Institute for Basic Science of Korea (grant no. IBS-R009-D1). The Excimer Laser COMPexPro 201F (Coherent Co.) for thin film growth and M-2000 ellipsometer (J.A.Woolam Co.) for optical measurements are supported by IBS Center for Correlated Electron Systems, Seoul National University.

**Author contributions**

B. K. and M. P. contributed equally to this work. B. K. and M. P. synthesized and characterized thin films. S. S., S. N., D. C., M. K., and M. K. performed and assisted in the magnetic property experiments. E. K. K. supported optical spectroscopy. B. K., M. P., C. S., J. M. O., and C. S. analyzed the experimental data. B. K. and C. S wrote the paper with input from all co-authors.

**Competing interests**

The authors declare no competing interests

[*]chsohn@unist.ac.kr
[**]okjongmok@pusan.ac.kr

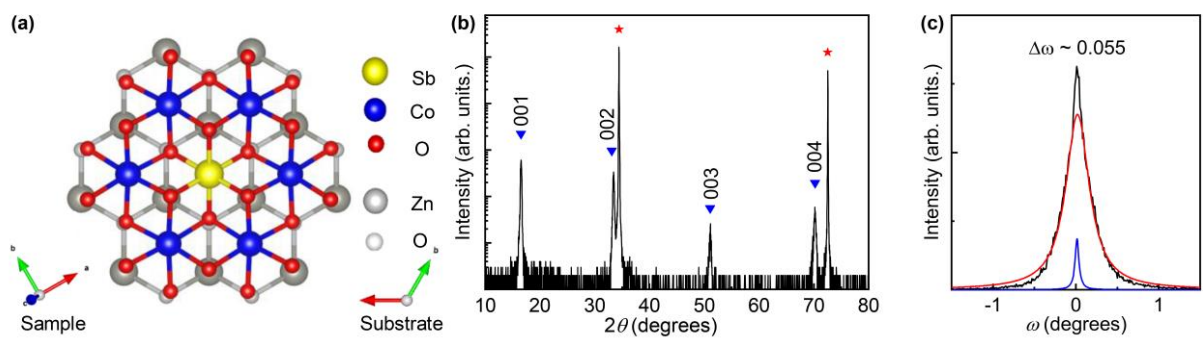

Figure 1. (a) Schematics of the epitaxial relationship between Na$_3$Co$_2$SbO$_6$ (001) and ZnO (0001) atomic planes. (b) XRD $\theta$-$2\theta$ data of Na$_3$Co$_2$SbO$_6$ thin film grown on ZnO [0001] substrate. The blue triangles and red stars indicate the 00$l$ peak of Na$_3$Co$_2$SbO$_6$ and the 000$l$ peak of ZnO, respectively. (c) Rocking curve of 001 peak of Na$_3$Co$_2$SbO$_6$ film. The blue and red lines are the fitting functions of two Lorentz oscillators.

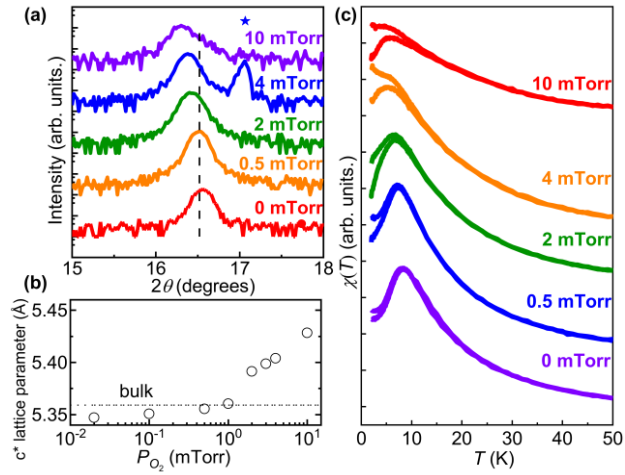

Figure 2. (a) The $P_{O_2}$ dependence of 001 peak of Na$_3$Co$_2$SbO$_6$ thin film. The dashed line is the 001 peak of bulk Na$_3$Co$_2$SbO$_6$. The blue star is the forbidden 0001 peak of the ZnO capping layer. (b) The $P_{O_2}$ dependence of $c^*$ lattice parameter. The dashed line indicates bulk one. (c) The $P_{O_2}$ dependence of ZFC/FC $\chi(T)$ of Na$_3$Co$_2$SbO$_6$.

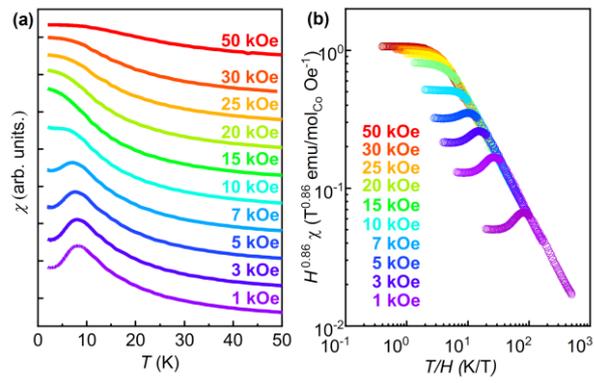

Figure 3. (a) Temperature-dependent in-plane $\chi(T)$ of Na$_3$Co$_2$SbO$_6$ heterostructure at various applied $H$. (b) Scaling of $H^{0.86}\chi(T)$ as a function $T/H$ on a log-log scale up to $H$ = 50 kOe.

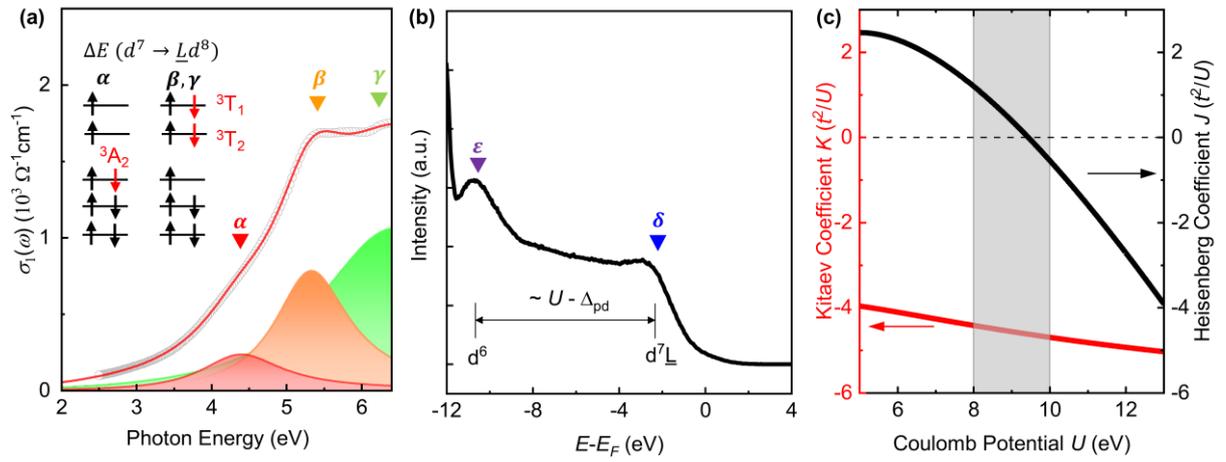

Figure 4. (a) The real part of optical conductivity $\sigma_1(\omega)$ of Na$_3$Co$_2$SbO$_6$. Open circles and solid lines are experimental data and fitting lines with Lorentz oscillators, respectively. The inset shows the schematic of possible charge-transfer transitions of $t_{2g}^5 e_g^2$ electron configuration. The red arrows represent excited electrons from O 2$p$ orbitals. (b) The valence band photoelectron spectrum of Na$_3$Co$_2$SbO$_6$. (c) Calculated the Heisenberg coefficient $J$ and the Kitaev coefficient $K$, in the units of $t^2/U$. The range of estimated $U$ is represented with a gray shaded area.

**Uncategorized References**

Na3Co2SbO6 and Na2Co2TeO6. *Journal of Physics: Condensed Matter* **34**, 045802 (2021).

19  Hwang, K., Go, A., Seong, J. H., Shibauchi, T. & Moon, E.-G. Identification of a Kitaev quantum spin liquid by magnetic field angle dependence. *Nature communications* **13**, 1-9 (2022).

20  Liu, X. *et al.* Proximate quantum spin liquid on designer lattice. *Nano letters* **21**, 2010-2017 (2021).

21  Yan, J.-Q. *et al.* Magnetic order in single crystals of Na 3 Co 2 SbO 6 with a honeycomb arrangement of 3 d 7 Co 2+ ions. *Physical Review Materials* **3**, 074405 (2019).

22  Csiszar, S. *et al.* Controlling orbital moment and spin orientation in CoO layers by strain. *Physical Review Letters* **95**, 187205 (2005).

23  Guo, E.-J. *et al.* Switchable orbital polarization and magnetization in strained LaCo O 3 films. *Physical Review Materials* **3**, 014407 (2019).

24  Kroemer, H. Quasi-electric and quasi-magnetic fields in nonuniform semiconductors. *RCA review* **18**, 332-342 (1957).

25  Baibich, M. N. *et al.* Giant magnetoresistance of (001) Fe/(001) Cr magnetic superlattices. *Physical review letters* **61**, 2472 (1988).

26  Minakawa, T., Murakami, Y., Koga, A. & Nasu, J. Majorana-mediated spin transport in Kitaev quantum spin liquids. *Physical Review Letters* **125**, 047204 (2020).

27  Jenderka, M. *et al.* Mott variable-range hopping and weak antilocalization effect in heteroepitaxial Na 2 IrO 3 thin films. *Physical Review B* **88**, 045111 (2013).

28  Jenderka, M., Schmidt-Grund, R., Grundmann, M. & Lorenz, M. Electronic excitations and structure of Li2IrO3 thin films grown on ZrO2: Y (001) substrates. *Journal of Applied Physics* **117**, 025304 (2015).

29  Yamashita, M., Gouchi, J., Uwatoko, Y., Kurita, N. & Tanaka, H. Sample dependence of half-integer quantized thermal Hall effect in the Kitaev spin-liquid candidate α− RuCl 3. *Physical Review B* **102**, 220404 (2020).

30  Viciu, L. *et al.* Structure and basic magnetic properties of the honeycomb lattice compounds Na2Co2TeO6 and Na3Co2SbO6. *Journal of Solid State Chemistry* **180**, 1060-1067 (2007).

31  Wong, C., Avdeev, M. & Ling, C. D. Zig-zag magnetic ordering in honeycomb-layered Na3Co2SbO6. *Journal of Solid State Chemistry* **243**, 18-22 (2016).

32  Stratan, M. I. *et al.* Synthesis, structure and magnetic properties of honeycomb-layered Li 3 Co 2 SbO 6 with new data on its sodium precursor, Na 3 Co 2 SbO 6. *New Journal of Chemistry* **43**, 13545-13553 (2019).

33  Liu, H. & Khaliullin, G. Pseudospin exchange interactions in d 7 cobalt compounds: possible realization of the Kitaev model. *Physical Review B* **97**, 014407 (2018).

34  Liu, H., Chaloupka, J. & Khaliullin, G. Kitaev spin liquid in 3 d transition metal compounds. *Physical Review Letters* **125**, 047201 (2020).

35  Li, X. *et al.* Giant Magnetic In-Plane Anisotropy and Competing Instabilities in
18

# Supplementary Information of

# Honeycomb oxide heterostructure: a new platform for Kitaev quantum spin liquid

Baekjune Kang,[1,†] Miju Park,[1,‡] Sehwan Song,[2] Seunghyun Noh,[3] Daeseong Choe,[3] Minsik Kong,[2] Minjae Kim,[2] Choongwon Seo,[1] Eun Kyo Ko,[4,5] Gangsan Yi,[1] Jung-woo Yoo,[3] Sungkyun Park,[2] Jong Mok Ok,[2,**] and Changhee Sohn[1,*]

[1] *Department of Physics, Ulsan National Institute of Science and Technology, Ulsan, 44919, Republic of Korea*

[2] *Department of Physics, Pusan National University, Pusan, 46241, Republic of Korea*

[3] *Department of Materials Science and Engineering, Ulsan National Institute of Science and Technology, Ulsan, 44919, Republic of Korea*

[4] *Center for Correlated Electron Systems, Institute for Basic Science (IBS), Seoul, 08826, Republic of Korea*

[5] *Department of Physics and Astronomy, Seoul National University, Seoul, 08826, Republic of Korea*

**Synthesis of $Na_3Co_2SbO_6$ thin film**

The crystal structure of $Na_3Co_2SbO_6$ is monoclinic layered oxides ($a$ = 5.3707 Å, $b$ = 9.2891 Å, $c$ = 5.6533 Å, $\beta$ = 108.566°, space group (C/2m, 12)), where $Na^+$ and $(Co_{2/3}Sb_{1/3}O_2)^-$ layers are alternatively stacked.[30] The $Co^{2+}$ ions construct the honeycomb lattice and the $Sb^{5+}$ ion is located at the center of the honeycomb.

We followed the previously reported synthesis of polycrystalline $Na_3Co_2SbO_6$ powder by solid-state reaction method[30]. We synthesize $Na_3Co_2SbO_6$ on another substrate $MgAl_2O_4$ [111]. The optimized growth condition on $MgAl_2O_4$ and $T$ = 625 °C, $P_{O_2}$ = 2 mTorr, $E$ = 0.9 J/cm$^2$, and laser repetition = 10 Hz. Figure S2a is the schematic of the interface between the $Na_3Co_2SbO_6$ (001) and $MgAl_2O_4$ (111) plane. Figure S2b is the XRD $\theta$-$2\theta$ data of $Na_3Co_2SbO_6$ thin film grown on $MgAl_2O_4$ [111] substrate. Figure S2c is the rocking curve of the 001 peak of $Na_3Co_2SbO_6$. The blue and red lines are the fitting function of two Lorentz oscillators. Figure S2d is the $\varphi$-scan of $Na_3Co_2SbO_6$ thin film on $MgAl_2O_4$ proving epitaxial growth.

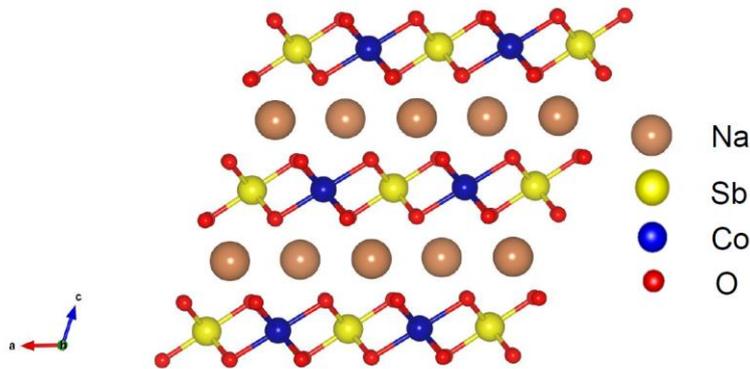

**Figure S1. The schematic of the lattice structure of $Na_3Co_2SbO_6$**

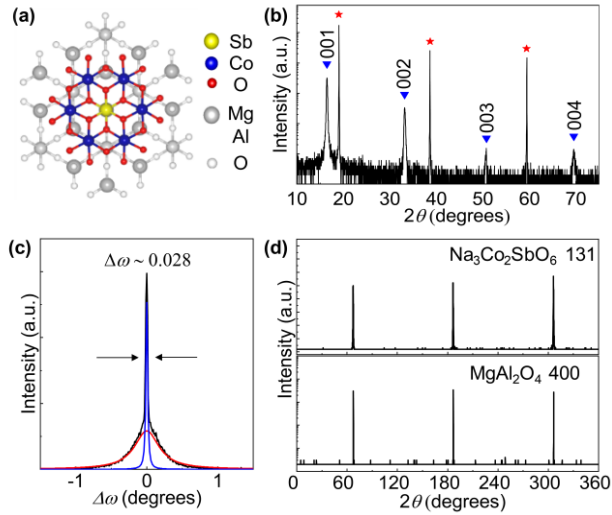

**Figure S2.** (a) Schematics of the epitaxial relationship between $Na_3Co_2SbO_6$ (001) and $MgAl_2O_4$ (111) atomic planes. (b) XRD $\theta$-$2\theta$ data of $Na_3Co_2SbO_6$ thin film grown on $MgAl_2O_4$ [111] substrate. The blue triangles and red stars indicate 00$l$ peaks of $Na_3Co_2SbO_6$ and $lll$ peaks of $MgAl_2O_4$, respectively. (c) Rocking curve of 001 peak of $Na_3Co_2SbO_6$ film. The blue and red lines are the fitting functions of two Lorentz oscillators. (d) The XRD azimuthal $\varphi$-scan with (131) plane of $Na_3Co_2SbO_6$ and (004) plane of $MgAl_2O_4$, respectively.

**The magnetic susceptibility measurement on thin films**

The bare data includes the sum of the magnetic response from $Na_3Co_2SbO_6$ film and ZnO substrate (Black line of figure S3). The magnetic susceptibility of $Na_3Co_2SbO_6$ (Blue line of figure S3) is obtained by deducting the summation from the substrate signals after normalizing mass (Red line of figure S3). A figure S4 is the contour plot of field-induced $\chi(T)$ which shows the three ground states, an antiferromagnetic state ($H <$ 10 kOe), an intermediate state (10 kOe $< H <$ 20 kOe), a fully polarized state (20 kOe $< H$). The intermediate state is recently elucidated as a ferrimagnetic state by inelastic neutron scattering.[29] Figure S5 are the $M$-$H$ curve and its derivatives of $Na_3Co_2SbO_6$ at 2.5 K. Clear peak at 10 kOe indicates the spin-flop transition consistent with the previous result.[35]

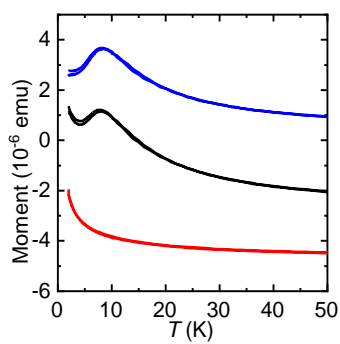

**Figure S3.** The $\chi(T)$ of ZnO and $Na_3Co_2SbO_6$. The black one is a mixed signal of ZnO and $Na_3Co_2SbO_6$. The red one is the magnetic susceptibility of ZnO alone. We extract $Na_3Co_2SbO_6$ magnetic susceptibility by deducing mass normalized response from the substrate.

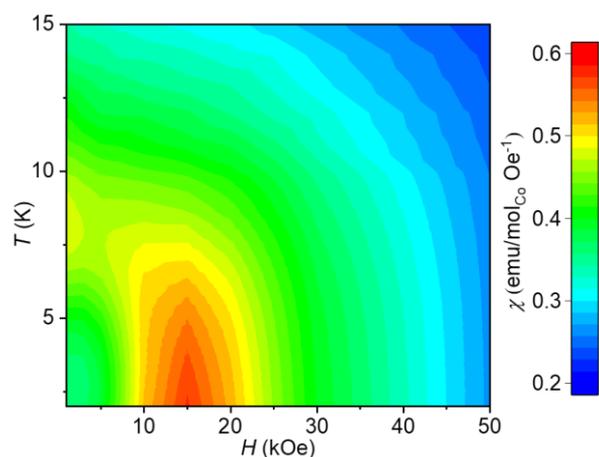

**Figure S4.** The contour plot of field-induced $\chi(T)$. The intermediate state ($H < 10$ kOe) is distinguished by an antiferromagnetic ground state (10 kOe $< H <$ 20 kOe) and a fully polarized state (20 kOe $< H$) with red color.

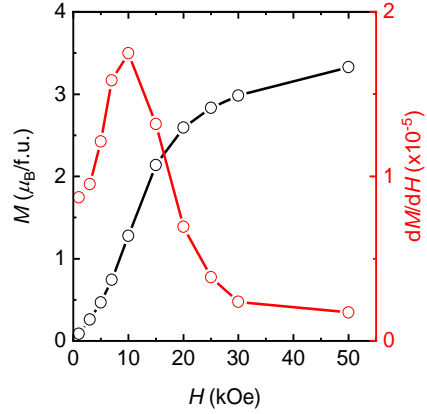

**Figure S5. The *M-H* curve and its derivatives of Na$_3$Co$_2$SbO$_6$ at 2.5 K. The derivatives graph shows a clear spin-flop transition at 10 kOe.**

**The electronic structure and optical *p-d* excitations of high spin $d^7$ cobalt ions**

We calculated the energies of optical excitation of $d^7$ high spin cobalt ion based on the multi-orbital Hubbard model. The low-energy excitations are transitions from oxygen 2*p* to cobalt $t_{2g}$ and $e_g$ orbitals. The final states consisting of two holes are equivalent to those consisting of two electrons due to electron-hole symmetry. We calculated the expected electrostatic energies of each wavefunction. After the electron is optically excited from oxygen 2*p* to cobalt 3*d* band, the wavefunctions constituted by two holes are described as[42]

$$|\Psi(ee^3A_2)(\bm{r_1},\bm{r_2})\rangle = |u_\uparrow(\bm{r_1})v_\uparrow(\bm{r_2})|$$

$$|\Psi(te^3T_1)(\bm{r_1},\bm{r_2})\rangle = |\xi_\uparrow(\bm{r_1})u_\uparrow(\bm{r_2})|$$

$$|\Psi(te^3T_2)(\bm{r_1},\bm{r_2})\rangle = |\xi_\uparrow(\bm{r_1})v_\uparrow(\bm{r_2})|$$

where $\xi_\uparrow = d_{xy}, d_{yz}, d_{zx}$, $u_\uparrow = d_{x^2-y^2}$, $v_\uparrow = d_{x^2-y^2}$ orbital with up-spin, | | is slater determinant satisfying the antisymmetric nature of fermion, *t/e* is a hole in $t_{2g}/e_g$ manifold, $^3A_2, ^3T_1, ^3T_2$ is a symmetry of wavefunction. All wavefunction is a triplet (*l* = 1), and there are three wavefunctions according to the magnetic quantum number. Due to the degeneracy of three wavefunctions, we choose the easiest one to compute. The expectation energy for the Coulomb Hamiltonian of each wavefunction is written as

$$\langle \Psi(ee^3A_2)|\frac{1}{r}|\Psi(ee^3A_2)\rangle = A - 8B$$

$$\langle \Psi(te^3T_2)|\frac{1}{r}|\Psi(te^3T_2)\rangle = A - 8B$$

$$\langle \Psi(te^3T_1)|\frac{1}{r}|\Psi(te^3T_1)\rangle = A + 4B$$

where $A$, $B$, and $C$ are called the Racah parameter. The Racah parameter is the function of slater integrals $F^k$

$$F^k = \int_0^\infty r_1^2 dr_1 \int_0^\infty r_2^2 dr_2\, R^2(r_1)R^2(r_2)\frac{r_<^k}{r_>^{k+1}}$$

where $R(r)$ is normalized radial part of the 3d orbital, $r_< = \min\{r_1, r_2\}$, $r_< = \max\{r_1, r_2\}$. The Slater integral is simplified as $F_0 = F^0$, $F_2 = F^2/49$, $F_4 = F^4/441$. The Racah parameter is written as

$$A = F_0 - 49F_4, B = F_2 - 5F_4, C = 35F_4$$

For the 3d electron, $5B \approx C$ satisfies.[33] The Hund's coupling is also written as the Racah parameter, $J_H = 3B + C \approx 8B$. Consequently, the energies of optical excitation are expressed as

$$E(^3A_2) = \Delta_{pd}, E(^3T_2) = \Delta_{pd} + 10Dq, E(^3T_1) = \Delta_{pd} + 10Dq + \frac{3}{2}J_H$$

, where $\Delta_{pd}$ is charge-transfer energy.

**The optical conductivity of Na$_3$Co$_2$SbO$_6$ on ZnO**

Figure S6a is the real part of the optical conductivity $\sigma_1(\omega)$ of ZnO. The peak at 3.3 eV is the absorption of ZnO which hinders the extracting accurate $\sigma_1(\omega)$ of Na$_3$Co$_2$SbO$_6$. Figure

S6b is the $\sigma_1(\omega)$ of $Na_3Co_2SbO_6$. The blue star is the artificial kink due to the sharp absorption of ZnO. The peak positions of $Na_3Co_2SbO_6$ on ZnO are consistent with the peak position on another $MgAl_2O_4$ substrate.

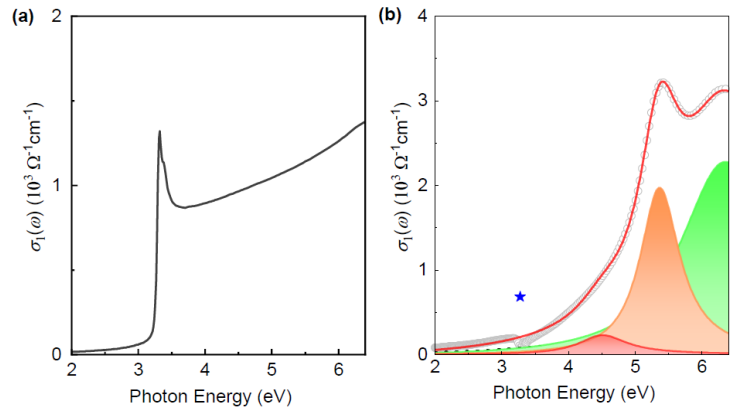

**Figure S6.** (a) The optical conductivity $\sigma_1(\omega)$ of ZnO substrate. (b) The optical conductivity $\sigma_1(\omega)$ of $Na_3Co_2SbO_6$ on ZnO substrate. The blue star is the artificial kink due to the sharp absorption peak of ZnO.